\documentclass[twocolumn,prb]{revtex4-2}
\usepackage{amsfonts}
\usepackage[T1]{fontenc}
\usepackage{amsmath,amsbsy,amssymb,graphicx}
\usepackage{times}

\begin{document}

\title{Dynamical nonlinear higher-order non-Hermitian skin effects \\
and topological trap-skin phase}
\author{Motohiko Ezawa}
\affiliation{Department of Applied Physics, University of Tokyo, Hongo 7-3-1, 113-8656,
Japan}

\begin{abstract}
We study nonreciprocal nonlinear Schr\"{o}dinger systems. As a prototype we
analyze the Hatano-Nelson model together with a typical nonlinear term
introduced and its generalization to two dimensions. We employ the quench
dynamics, where a pulse is given to one site and its time evolution is
analyzed. It is found that the skin state is always formed due to the
nonreciprocal hopping and hence that the system is topological. However, the
structure of the skin state is essentially modified by the nonlinear
interaction because it favors a self-trapping. Four typically different
states emerge as an interplay between these two interactions, depending how
the pulse is trapped to the initial site. They are the skin, trap-skin,
shifted-trap-skin and embedded-trap-skin states, forming four phases in the
one-dimensional model. The phase boundary is determined by a gap in terms of
certain phase indicators. On the other hand, we find three phases with the
shifted-trap-skin phase being absent in the two-dimensional model.
\end{abstract}

\maketitle

\section{Introduction}

Non-Hermitian topological physics is one of the most exciting fields of
current condensed matter physics\cite%
{Bender,Bender2,Kohmoto,Scho,Weimann,Liang,Pan,Nori,Zhu,Konotop,Fu,UedaPRX,Gana,Katsura,Yuce,LangWang,UedaHOTI,EzawaLCR,EzawaSkin}%
. Among them, the skin state is prominent\cite%
{Xiong,Mart,UedaPRX,Kunst,Yao,Lee,Jin,Research,SkinTop,Luo,EzawaSkin}, which
is absent in the Hermitian systems. All the states are localized at one edge
in the skin state for a finite chain. The skin state is generated in a
nonreciprocal hopping model, where the amplitudes of the right-going and
left-going hoppings are different. It is generalized to higher-order skin
states in higher-dimensional systems\cite%
{SkinTop,EzawaSkin,KawabataSkin,Okugawa}. On the other hand, nonlinear
physics attracts renewed interests in the context of topological physics. It
is studied in mechanical\cite{Snee,PWLo,MechaRot}, photonic\cite%
{Ley,Zhou,MacZ,Smi,Tulo,Kruk,NLPhoto,Kirch,Sin,Laser}, electric circuit\cite%
{Hadad,Sone,TopoToda} and resonator\cite{Zange} systems.

It is interesting and important to explore systems in which nonreciprocity
and nonlinearity coexist to reveal phenomena ascribed to their competition
and collaboration. Photonic system provides us with a typical playground\cite%
{KhaniPhoto,Hafe2,Hafezi,WuHu,TopoPhoto,Ozawa16,Ley,KhaniSh,Zhou,Jean,Ota18,Ozawa,Ota19,OzawaR,Hassan,Ota,Li,Yoshimi,Kim,Iwamoto21}%
, where the nonreciprocal hopping is realized based on coupled resonant
optical waveguides\cite{ZuNR,SongNR} and the nonlinearity is introduced by
the Kerr effect\cite{Szameit,Chris}.

In this paper, as a prototype of such systems, we analyze the Hatano-Nelson
model\cite{Hatano} modified to include a typical nonlinear term. We also
analyzed its generalization to two dimensions. {It is an intriguing
feature of the models that the topological number of the skin state is given
solely by the direction of the localization whether the system is linear or
nonlinear. Hence, the skin state is topologically robust how much it is
distorted as long as the rightward or leftward localization occurs. 

To investigate a rich physics induced by the nonlinear effects, we
study the quench dynamics by giving a pulse to one site and exploring the
time evolution of the classical field $\psi _{n}(t)$ subject to the discrete
nonreciprocal nonlinear Schr\"{o}dinger equation. First, we consider the
one-dimensional model. When the nonlinearity is small enough, on the other
hand, the initial pulse is pushed toward the right edge, forming a
well-known skin state, as in the linear model. When it is strong enough, the
pulse is trapped to the initial site together with a formation of the skin
state. We call it the trap-skin state. Novel phenomena occur as an interplay
between nonreciprocity and nonlinearity in the intermediate regime. We have
found a state where the pulse is trapped not to the initial site but to the
adjacent site, which we call the shifted-trap-skin state. We have also found
a state where the pulse is trapped to the initial site although it is almost
embedded in the skin state. We call it the embedded-trap-skin state. These
four different patterns of states occupy some continuous regions in the $%
\lambda $-$\xi $ space with the nonreciprocity strength parameter $\lambda $
and the nonlinearity strength parameter $\xi $. We have thus found four
phases, the skin phase, the shifted-trap-skin phase, the embedded-trap-skin
phase and the trap-skin phase in the one-dimensional model. The phase
indicators are $S_{n}$ which are the time average of the amplitude $|\psi
_{n}(t)|$, where $n=m_{0}$ and $m_{0}+1$, with $m_{0}$ representing the site
to which the pulse is given. The phase boundary is determined by the
position of a gap in these phase indicators. On the other hand, we have
found that the shifted-trap-skin phase is absent in the two-dimensional
model.

\section{One-dimensional model}

\subsection{Nonlinear skin effect}

We propose a one-dimensional discrete nonreciprocal nonlinear Schr\"{o}%
dinger equation defined by%
\begin{equation}
i\frac{d\psi _{n}}{dt}+\kappa _{\text{R}}\psi _{n+1}+\kappa _{\text{L}}\psi
_{n-1}-\left( \kappa _{\text{R}}+\kappa _{\text{L}}\right) \psi _{n}+\xi
\left\vert \psi _{n}\right\vert ^{2}\psi _{n}=0,  \label{DS}
\end{equation}%
where $\kappa _{\text{R}}$ and $\kappa _{\text{L}}$ are the right-going and
left-going hopping amplitudes, respectively. It is constructed by
introducing a typical nonlinear term to the Hatano-Nelson model\cite{Hatano}%
. The system is nonreciprocal if $\kappa _{\text{R}}\neq \kappa _{\text{L}}$%
. The nonlinearity is controlled by the parameter $\xi $, where large $\xi $
indicates strong nonlinearity.

\begin{figure*}[t]
\centerline{\includegraphics[width=0.98\textwidth]{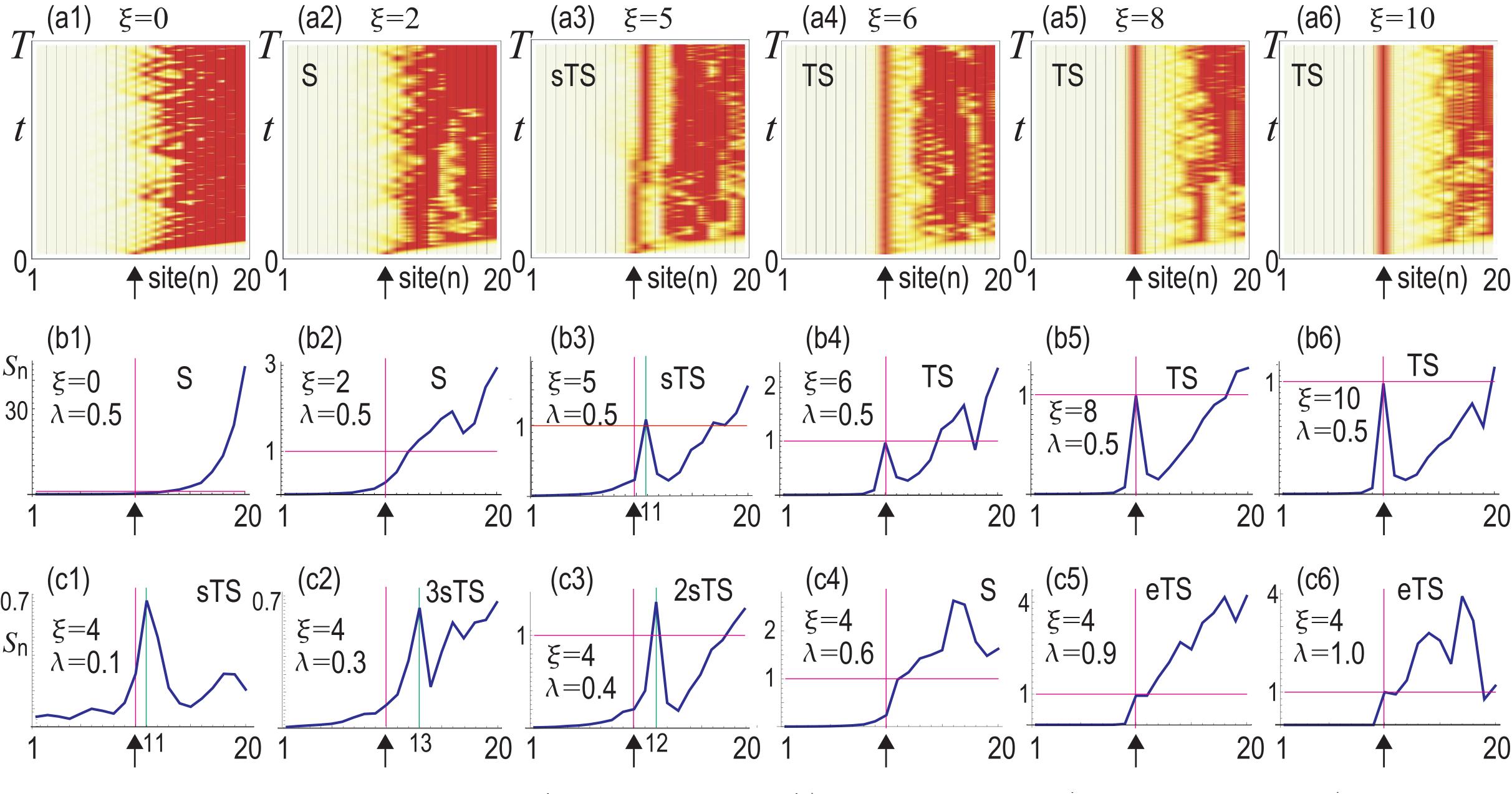}}
\caption{(a1)$\sim $(a6) Time evolution of the amplitude $|\protect\psi %
_{n}(t)| $ from $t=0$ to $t=T$ for various nonlinearity $\protect\xi $ at
fixed nonreciprocity $\protect\lambda =0.5$. The horizontal axis is the site
index $n$, and the vertical axis is time $t$. The color density indicates
the magnitude of $|\protect\psi _{n}(t)|$. (b1)$\sim $(b6) Time average $%
S_{n}$ of the amplitude defined by Eq.(\protect\ref{Sn}) for various $%
\protect\xi $ at fixed $\protect\lambda =0.5$. (c1)$\sim $(c6) Time average $%
S_{n}$ for various $\protect\lambda $ at fixed $\protect\xi =4$. The
horizontal axis is the site index $n$, and the vertical axis is the time
average $S_{n}$. The arrow indicates the initial site $m_0=L/2=10$ from
which the quench dynamics starts. We have set $\protect\kappa =1$, $L=20$
and $T=50 $ in units of $1/\protect\kappa$. Symbols S, TS, sTS, eTS stand
for the skin, trap-skin, shifted-trap-skin and embedded-trap-skin states,
respectively.}
\label{FigSkinDensity}
\end{figure*}

We parametrize%
\begin{equation}
\kappa _{\text{R}}=\kappa \left( 1+\lambda \right) ,\qquad \kappa _{\text{L}%
}=\kappa \left( 1-\lambda \right) ,  \label{kappa}
\end{equation}%
where $\lambda $ is referred to as the nonreciprocity, $-1<\lambda <1$.

\subsection{Quench dynamics}

Quench dynamics provides us with a good signal to detect various phases\cite%
{QWalk}, which is applicable to nonlinear systems\cite%
{TopoToda,MechaRot,NLPhoto,Sin} as well. Let us study a quench dynamics
starting from the site indexed by $m_{0}$,%
\begin{equation}
\psi _{n}\left( t\right) =\delta _{n,m_{0}}\quad \text{at}\quad t=0.
\label{IniCon}
\end{equation}%
Here, we take the site $m_{0}$ at the center of the sample, i.e., $m_{0}=L/2$%
\ for a finite chain with size $L$, and $m_{0}=(L/2,L/2)$ for a finite
square with size $L\times L$, where $L$ is an even integer. Namely, solving
the dynamical equation (\ref{DS}) under the initial condition (\ref{IniCon}%
), we study the time evolution $\psi _{n}\left( t\right) $ as a function of
the system parameters $\xi $ and $\lambda $.

\subsection{Linear model}

The Hatano-Nelson Hamiltonian is given by\cite{Hatano}%
\begin{equation}
H_{nm}=\kappa _{\text{R}}\delta _{n,m+1}+\kappa _{\text{L}}\delta
_{n,m-1}-(\kappa _{\text{R}}+\kappa _{\text{L}})\delta _{n,m}.
\label{HataNel}
\end{equation}%
The eigenenergy is given in the momentum space by%
\begin{equation}
E\left( k\right) =\kappa _{\text{R}}e^{ik}+\kappa _{\text{L}}e^{-ik}-(\kappa
_{\text{R}}+\kappa _{\text{L}}),  \label{OneBnad}
\end{equation}%
which is complex for $\kappa _{\text{R}}\neq \kappa _{\text{L}}$. The system
is analytically solvable.

In the case of a finite chain, the skin state is generated at the left edge
and given by%
\begin{equation}
\psi _{2n+1}=\left( -\frac{\kappa _{\text{R}}}{\kappa _{\text{L}}}\right)
^{n}\psi _{1},\qquad \psi _{2n}=0
\end{equation}%
for $\left\vert \kappa _{\text{L}}\right\vert >\left\vert \kappa _{\text{R}%
}\right\vert $, while it is generated at the right edge and given by%
\begin{equation}
\psi _{L-2n}=\left( -\frac{\kappa _{\text{L}}}{\kappa _{\text{R}}}\right)
^{n}\psi _{L},\qquad \psi _{2n-1}=0.  \label{SkinR}
\end{equation}%
for $\left\vert \kappa _{\text{L}}\right\vert <\left\vert \kappa _{\text{R}%
}\right\vert $.

The dynamical Hatano-Nelson model is defined by%
\begin{equation}
i\frac{d\psi _{n}}{dt}+H_{nm}\psi _{m}=0,  \label{dynaHataNel}
\end{equation}%
which is Eq.(\ref{DS}) without the nonlinear term. The quench dynamics of
this linear model is exactly solvable in the case of an infinite chain.
Indeed, an analytic solution of the quench dynamics based on Eq.(\ref%
{dynaHataNel}) is constructed as%
\begin{equation}
\psi _{n}\left( t\right) =\left( i\sqrt{\frac{\kappa _{\text{L}}}{\kappa _{%
\text{R}}}}\right) ^{n-m_{0}}J_{\left\vert n-m_{0}\right\vert }\left( 2\sqrt{%
\kappa _{\text{L}}\kappa _{\text{R}}}t\right) ,  \label{EqA}
\end{equation}%
where $m_{0}$ represents the initial site as in Eq.(\ref{IniCon}) and 
$J_{\left\vert n-m_{0}\right\vert }$ is the Bessel function of the first
kind. It is straightforward to check that Eq.(\ref{EqA}) satisfies the
dynamical Hatano-Nelson model (\ref{dynaHataNel}) together with the initial
condition (\ref{IniCon}) with the aid of the formula%
\begin{equation}
\frac{d}{dt}\Psi _{n}\left( at\right) =\frac{ia}{2}\left[ \gamma \Psi
_{n-1}\left( at\right) +\frac{1}{\gamma }\Psi _{n+1}\left( at\right) \right]
,
\end{equation}%
where $\Psi _{n}\left( t\right) \equiv \left( i\gamma \right)
^{n}J_{n}\left( at\right) $ with an arbitrary constant $a$.

\subsection{Topological number}

\label{SecTopo}

The Hatano-Nelson model has only one band as in Eq.(\ref{OneBnad}) and there
is no "gap" in the usual acceptation of the Hermitian model. Nevertheless,
it is possible to define the topological number with the aid of the complex
degrees of freedom present in the energy\cite{Fu}. Indeed, the structure of
the band (\ref{OneBnad}) is an ellipse in the complex plane as illustrated
in Fig.\ref{FigHatanoRing}. Such a band is said to have a point gap\cite%
{UedaPRX,KawabataPRX}. The following observation is essential. As the
momentum runs from $k=0$ to $2\pi $, the band energy $E(k)$ encircles the
ellipse once. Furthermore, the direction of the winding is opposite for $%
\lambda >0$ and $\lambda <0$, or $\left\vert \kappa _{\text{L}}\right\vert
<\left\vert \kappa _{\text{R}}\right\vert $ and $\left\vert \kappa _{\text{L}%
}\right\vert >\left\vert \kappa _{\text{R}}\right\vert $. This property
allows us to define the topological number.

The topological number is defined in the Hatano-Nelson model by the winding
number in the complex-energy plane as\cite{Fu,UedaPRX}%
\begin{equation}
w=\int_{0}^{2\pi }\frac{dk}{2\pi }\partial _{k}\ln \left[ E\left( \mathbf{k}%
\right) -\bar{E}\right] ,  \label{wind}
\end{equation}%
where $\bar{E}$ is the mean energy,%
\begin{equation}
\bar{E}=\int_{0}^{2\pi }\frac{dk}{2\pi }E\left( \mathbf{k}\right) =-(\kappa
_{\text{R}}+\kappa _{\text{L}}).
\end{equation}%
It yields $w=1$ for $\left\vert \kappa _{\text{L}}\right\vert <\left\vert
\kappa _{\text{R}}\right\vert $ and $w=-1$ for $\left\vert \kappa _{\text{L}%
}\right\vert >\left\vert \kappa _{\text{R}}\right\vert $. Furthermore, we
have $w=0$ for $\kappa _{\text{L}}=\kappa _{\text{R}}$, because $E\left( 
\mathbf{k}\right) $ is single-valued. It is understood as the vorticity of
the energy\cite{Fu}.

\begin{figure}[t]
\centerline{\includegraphics[width=0.48\textwidth]{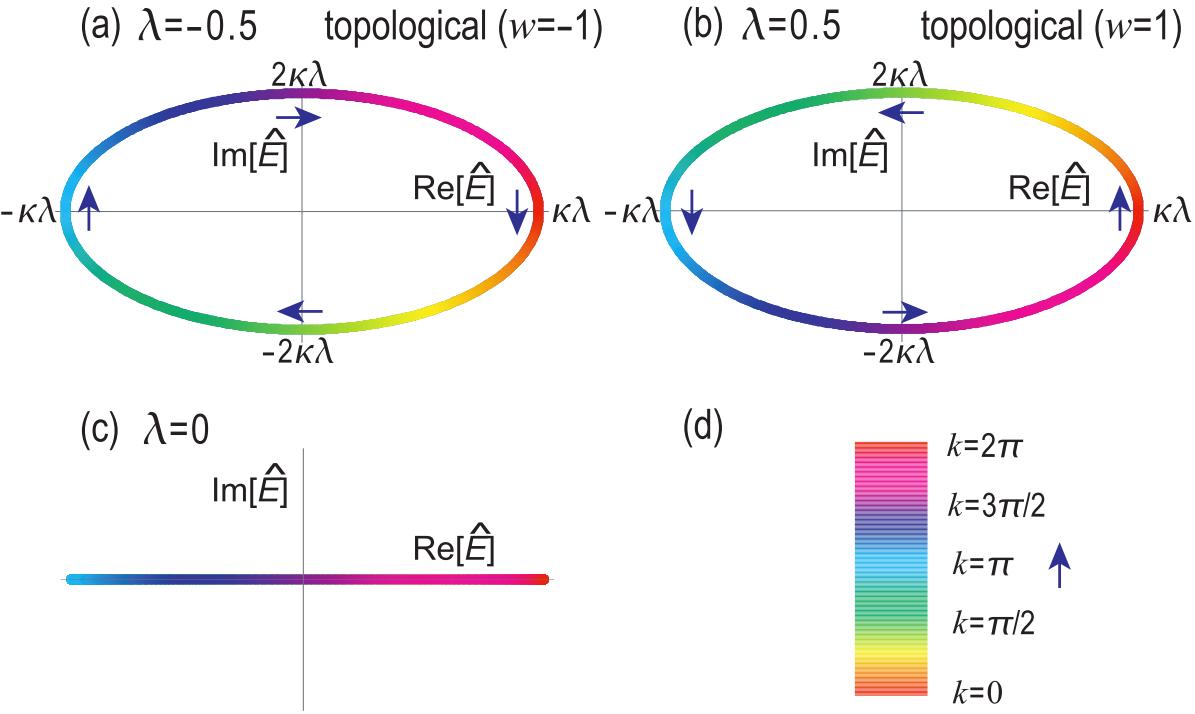}}
\caption{Band structure in the Re[$\hat{E}$]-Im[$\hat{E}$] plane, where $%
\hat{E}\equiv E\left( \mathbf{k}\right) -\bar{E}$ at (a) $\protect\lambda %
=-0.5$ with $w=-1$, (b) $\protect\lambda =0.5$ with $w=1$, and (c) $\protect%
\lambda =0$ with $w=0$. Here, $w$ is the winding number. (d) The color
palette indicates the momentum $k$ for (a), (b) and (c). The arrow indicates
the direction for $k$ to increase.}
\label{FigHatanoRing}
\end{figure}

The topological number is rewritten as%
\begin{equation}
w=\text{sgn}(\left\vert \psi _{L}\right\vert -\left\vert \psi
_{1}\right\vert ),  \label{sgn}
\end{equation}%
in terms of the amplitudes at the left and right edges of a finite chain.
This formula is valid even for nonlinear systems. The system is topological
for $\kappa _{\text{L}}\neq \kappa _{\text{R}}$ and trivial for $\kappa _{%
\text{L}}=\kappa _{\text{R}}$.

It is an intriguing feature of the Hatano-Nelson type models that the
topological number is determined solely by the direction of the localization
in the skin state based on the formula (\ref{sgn}). Namely, the skin state
is topologically robust how much it is distorted as long as the rightward or
leftward localization occurs. We will numerically check the robustness
against disorders in Subsection \ref{SecDisorder}.

\begin{figure*}[t]
\centerline{\includegraphics[width=0.98\textwidth]{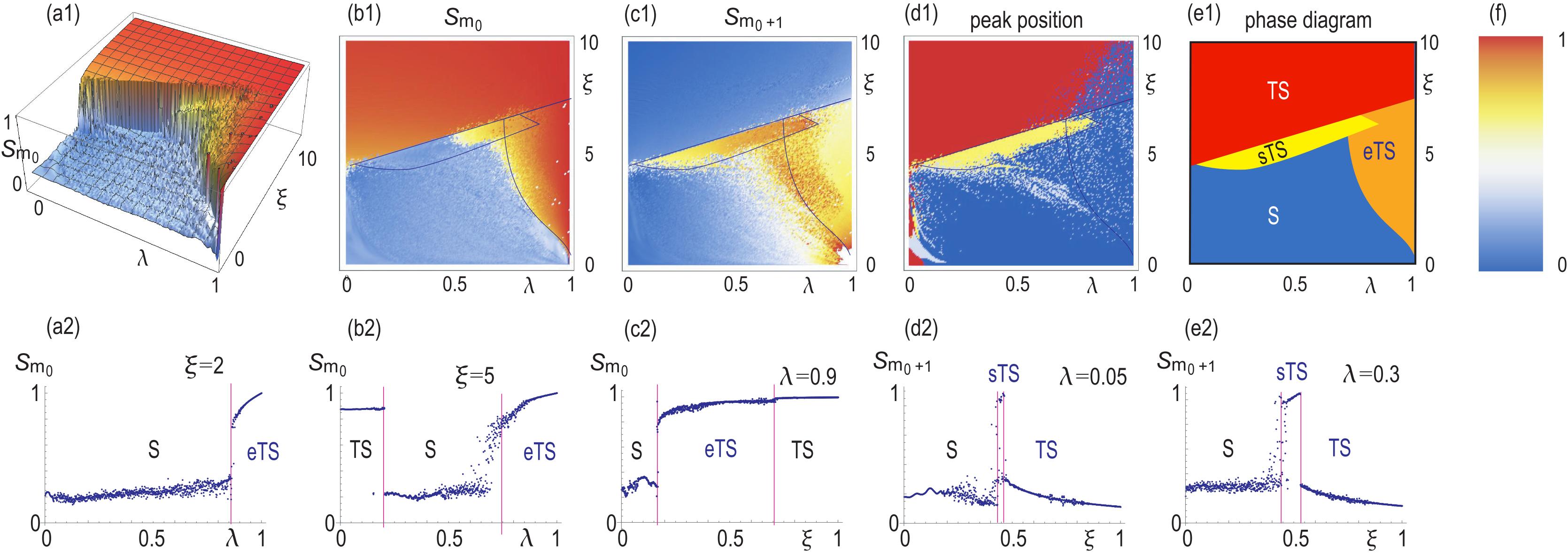}}
\caption{(a1) and (b1) Time average $S_{m_{0}}$ of the amplitude as a
function of the nonreciprocity $\protect\lambda $ and the nonlinearity $%
\protect\xi $ in the one-dimensional model, with (a1) being a bird's eye's
view. The red region represents $S_{m_{0}}=1$, while the blue region
represents $S_{m_{0}}=0$. (c1) The time average $S_{m_{0}+1}$ differentiates
the trap-skin (TS) phase from all others. The embedded-trap-skin (eTS) phase
is assigned from (b1) and (c1). (d1) Positions of the maximum among $%
S_{m_{0}},S_{m_{0}+1},S_{m_{0}+2}$ and $S_{m_{0}+3}$ are shown, where the
maximum is taken at $S_{m_{0}}$ in the red region, at $S_{m_{0}+1}$ in the
yellow region, at $S_{m_{0}+2}$ in the white region, and at $S_{m_{0}+3}$ in
the blue region. The shifted-trap-skin (sTS) phase is assigned from (c1) and
(d1). (e1) Phase diagram in the $\protect\lambda $-$\protect\xi $ plane. (a2)%
$\sim $(e2) Typical examples of the average $S_{m_{0}}$ or $S_{m_{0}+1}$ as
a function of $\protect\lambda $ at a fixed value of $\protect\xi $ or a
function of $\protect\xi $ at a fixed value of $\protect\lambda $. The phase
boundary is determined by the position of a gap. We have set $\protect\kappa %
=1$ and $L=20$. (f) The color palette indicates the amplitude in (a1),
(b1) and (c1).}
\label{FigPhase}
\end{figure*}

\subsection{Strong nonlinear model}

We next study the strong nonlinear model, where we neglect the hopping term
with respect to the nonlinear term. We approximate Eq.(\ref{DS}) as%
\begin{equation}
i\frac{d\psi _{n}}{dt}=-\xi \left\vert \psi _{n}\right\vert ^{2}\psi _{n},
\end{equation}%
where all equations are separated one another. This system is also
analytically solvable.

We set $\psi _{n}\left( t\right) =r_{n}e^{i\theta _{n}\left( t\right) }$,
and make an ansatz that $r_{n}$ is a constant in the time $t$. Then, the
solution is given by $\theta _{n}=\xi r_{n}^{2}t+c$. Due to the initial
conservation (\ref{IniCon}), we find 
\begin{equation}
|\psi _{n}\left( t\right) |=\delta _{nm_0}.  \label{TrapCond}
\end{equation}%
Hence, the initial pulse is trapped to the initial site $n=m$. This phase
may be referred to as the trap phase.

\subsection{Four different skin states}

We show the amplitude $|\psi _{n}|$ in Fig.\ref{FigSkinDensity} for various $%
\xi $ at fixed $\lambda =0.5$, and also for various $\lambda >0$ at fixed $%
\xi =4$, by numerically solving the quench dynamics subject to the equation
of motion (\ref{DS}) with the initial condition (\ref{IniCon}). The time
evolution is given in Figs.\ref{FigSkinDensity}(a1)$\sim $(a6) for each site 
$n$, where the value of $|\psi _{n}|$ is indicated by the darkness of color.

There is a characteristic behavior at the initial stage for each set of $\xi 
$ and $\lambda $. To remove it and to reveal a more quantitative structure
of the amplitude, we define the time average of its site dependence over a
time span from $T/2$ to $T$ by 
\begin{equation}
S_{n}\equiv \frac{1}{T/2}\int_{T/2}^{T}\left\vert \psi _{n}\left( t\right)
\right\vert dt,  \label{Sn}
\end{equation}%
and show it in Figs.\ref{FigSkinDensity}(b1)$\sim $(b6) and (c1)$\sim $(c6)
for each site $n$.

As a general structure, the wave packet shifts rightward (leftward) when the
right-going hopping is larger (smaller) than the left-going hopping, forming
a skin state irrespective of $\lambda $ and $\xi $. Hence, the system is
always in the skin phase and topological with $w=1$ for $\lambda >0$ or $%
\left\vert \kappa _{\text{R}}\right\vert >\left\vert \kappa _{\text{L}%
}\right\vert $, and with $w=-1$ for $\lambda <0$ or $\left\vert \kappa _{%
\text{R}}\right\vert <\left\vert \kappa _{\text{L}}\right\vert $ even in the
nonlinear system according to the formula (\ref{sgn}). We assume $\lambda >0$
in what follows.

There are typically four different patterns in the skin states generated by
the nonlinearity effects as described below:

(1) The standard skin states emerge as in Figs.\ref{FigSkinDensity}(b1)$\sim 
$(b2) and (c4). We call it the skin (S) state.

(2) In addition to the formation of the skin state, the initial pulse is
trapped to the initial site as in Figs.\ref{FigSkinDensity}(b4)$\sim $(b6).
Namely, it is a coexistent system of the skin state and the trap state. The
characteristic feature is that the equality $S_{m_0}=1$ keeps to hold as a
reminiscence of the initial condition (\ref{IniCon}). We call it the
trap-skin (TS) state.

(3) In addition to the formation of the trap-skin state, the initial pulse
is trapped not to the initial site but to a site near to it as in Figs.\ref%
{FigSkinDensity}(b3) and (c1)$\sim $(c3). Let us call it the $n$%
-site-shifted-trap-skin state, and abbreviate it as the $n$sTS state, when
the trap site is away from the original site by $n$ sites. When $n=1$, we
call it simply the sTS state.

(4) The skin state develops over a wide region including the initial state $%
m_0 $ as in Figs.\ref{FigSkinDensity}(c5)$\sim $(c6). Although it looks just
as if it were the standard skin state, the equality $S_{m_0}=1$ keeps to
hold as a reminiscence of the initial condition (\ref{IniCon}). We call it
the embedded-trap-skin (eTS) state.

\subsection{Phase diagram}

We have pointed out the emergence of four types of the skin states depending
on $\lambda $ and $\xi $. Let us construct a phase diagram in the $\lambda $-%
$\xi $ space. First, we focus on the time average $S_{m_{0}}$, and show it
as a function of $\lambda $ and $\xi $ in Figs.\ref{FigPhase}(a1) and (b1),
where Fig.\ref{FigPhase}(a1) is a bird's eye's view.

Fig.\ref{FigPhase}(b1) may present a rough picture of the phase diagram,
where the blue region is the skin phase while the red region is the trap
phase. However, it is impossible to differentiate the trap-skin and the
embedded-trap-skin states in Fig.\ref{FigPhase}(b1) because $S_{m_{0}}=1$\
for both states. Furthermore, the transition between the skin and trap
phases is blurred around $\lambda =0.6$ and $\xi =0.6$.

They are differentiated by the indicator $S_{m_{0}+1}$, whose results are
shown in Fig.\ref{FigPhase}(c1). Note that $S_{m_{0}+1}$ is almost zero for
the trap-skin state but takes a larger value for the shifted-trap-skin
state. The phase boundary is linear as a function of $\lambda $ in Fig.\ref%
{FigPhase}(c1). It is understood as follows. The hopping motion is enhanced
as a function of the nonreciprocity $\lambda $ as in Eq.(\ref{kappa}). On
the other hand, it is depressed as a function of the nonlinearity $\xi $\
because it favors the trap state.

Next, we plot the position of the peak among $S_{m_{0}}$, $S_{m_{0}+1}$, $%
S_{m_{0}+2}$ and $S_{m_{0}+3}$. If the peak position is at $S_{m_{0}}$, it
is the trap-skin state depicted in red. If it is at $S_{m_{0}+n}$, it is the 
$n$-shifted-trap-skin state depicted in yellow for $n=1$, in white for $n=2$%
, in blue for $n=3.$ We find one-site-shifted-trap-skin states form a region
in the vicinity of the trap-skin state. On the other hand, $n$%
-site-shifted-trap-skin states do not form a continuous region for $n=2$ and 
$3$. Note that $3$-site-shifted-trap-skin states are buried in the skin
phase and hardly identified.

We thus obtain a phase diagram as in Fig.\ref{FigPhase}(e1). In determining
phase boundaries, we have calculated $S_{m_{0}}$ and $S_{m_{0}+1}$ as a
function of $\lambda $ for a fixed value of $\xi $ as in Figs.\ref{FigPhase}%
(a2) and (b2), and also as a function of $\xi $ for a fixed value of $%
\lambda $ as in Figs.\ref{FigPhase}(c2), (d3) and (e2) for many fixed
values. Phase transition points are clearly observed as gaps in these
figures.

\begin{figure}[t]
\centerline{\includegraphics[width=0.48\textwidth]{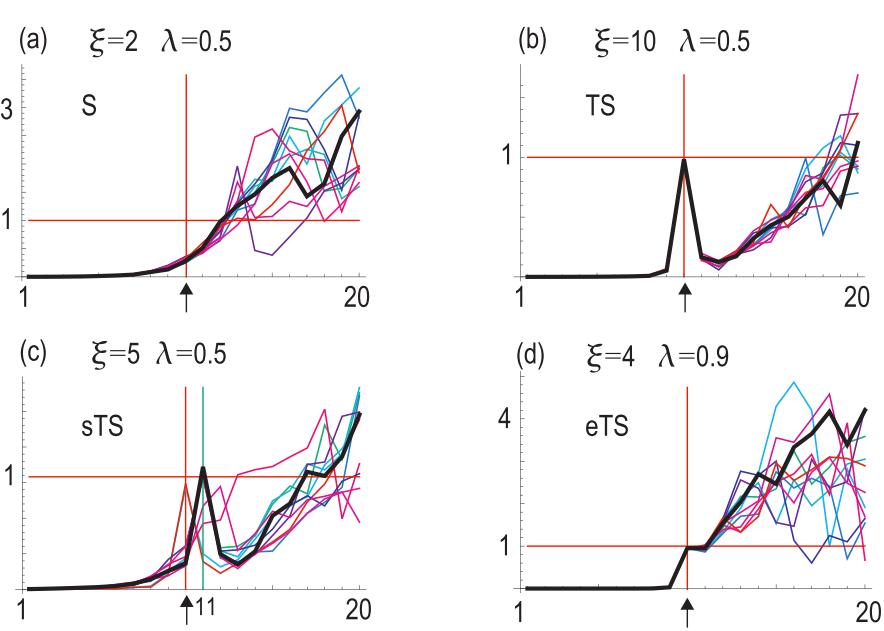}}
\caption{On-site disorder effects of the time average $S_{n}$ for a
typical state taken from Fig.\protect\ref{FigPhase} in each phase: (a) $%
\protect\xi =2,\protect\lambda =0.5$, (b) $\protect\xi =10,\protect\lambda %
=0.5$, (c) $\protect\xi =5,\protect\lambda =0.5$, (d) $\protect\xi =4,%
\protect\lambda =0.9$. Thick black curves are the amplitudes without
disorders. Colored curves are the amplitudes with disorders ranging from $%
\protect\zeta =0.1$ (red) to $\protect\zeta =1$ (cyan). We have set $\protect%
\kappa =1$ and $L=20 $. See also the caption of Fig.\protect\ref{FigPhase}.}
\label{FigDisorder}
\end{figure}

\begin{figure}[t]
\centerline{\includegraphics[width=0.48\textwidth]{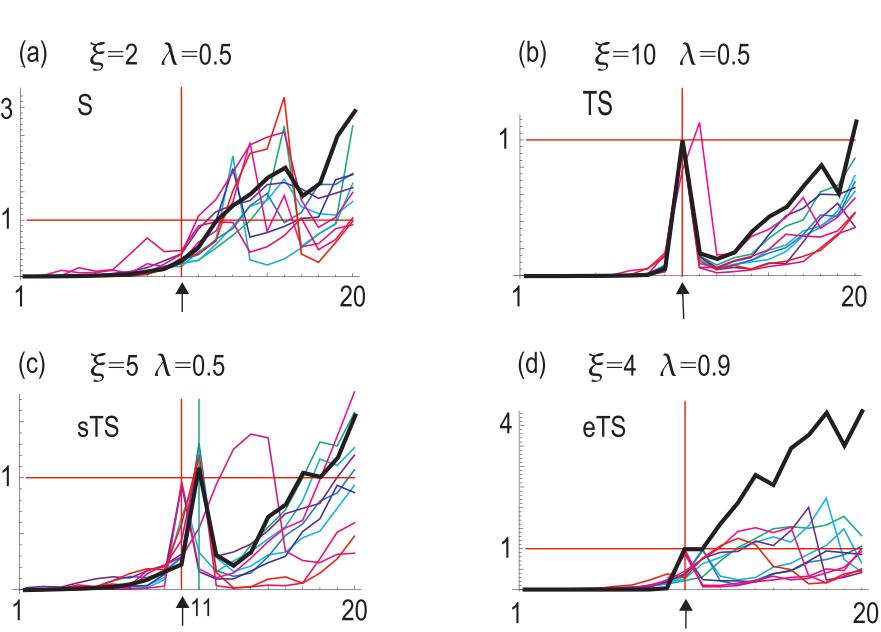}}
\caption{Hopping disorder effects of the time average $S_{n}$ for a
typical state taken from Fig.\protect\ref{FigPhase} in each phase: (a) $%
\protect\xi =2,\protect\lambda =0.5$, (b) $\protect\xi =10,\protect\lambda %
=0.5$, (c) $\protect\xi =5,\protect\lambda =0.5$, (d) $\protect\xi =4,%
\protect\lambda =0.9$. Thick black curves are the amplitudes without
disorders. Colored curves are the amplitudes with disorders ranging from $%
\protect\zeta =0.1$ (red) to $\protect\zeta =0.8$ (cyan). We have set $%
\protect\kappa =1$ and $L=20 $. See also the caption of Fig.\protect\ref%
{FigPhase}.}
\label{FigDisorderK}
\end{figure}

\begin{figure}[t]
\centerline{\includegraphics[width=0.48\textwidth]{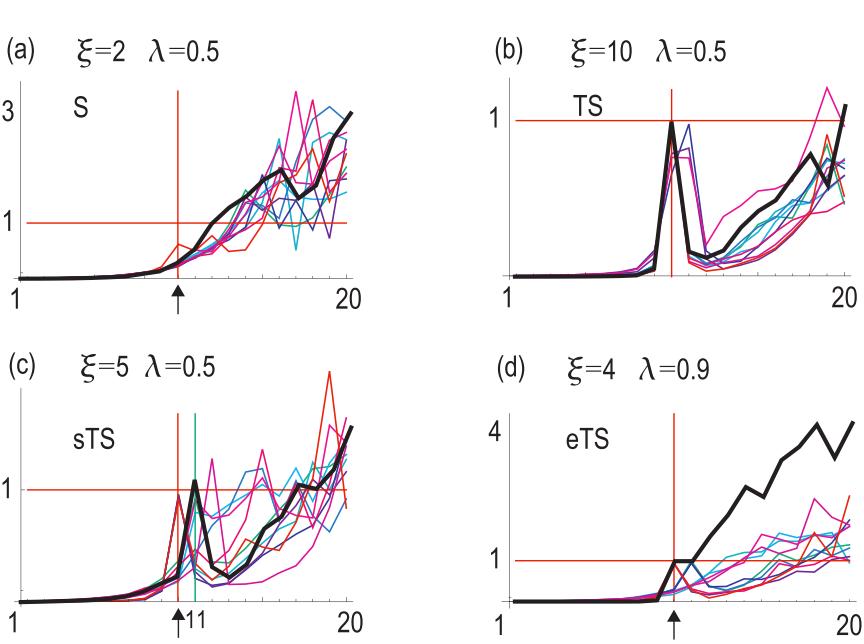}}
\caption{Nonlinearity disorder effects of the time average $S_{n}$
for a typical state taken from Fig.\protect\ref{FigPhase} in each phase: (a) 
$\protect\xi =2,\protect\lambda =0.5$, (b) $\protect\xi =10,\protect\lambda %
=0.5$, (c) $\protect\xi =5,\protect\lambda =0.5$, (d) $\protect\xi =4,%
\protect\lambda =0.9$. Thick black curves are the amplitudes without
disorders. Colored curves are the amplitudes with disorders ranging from $%
\protect\zeta =0.1$ (red) to $\protect\zeta =1$ (cyan). We have set $\protect%
\kappa =1$ and $L=20 $. See also the caption of Fig.\protect\ref{FigPhase}.}
\label{FigDisorderZ}
\end{figure}

\subsection{Disorder effects}

\label{SecDisorder}

As we have noted in Subsection \ref{SecTopo}, the robustness of the
topological states and that of the skin states are equivalent in the
Hatano-Nelson model. The topological states are well known to be robust
against disorders in the linear model. We study and confirm the robustness
of topological states in the nonlinear model for three types of disorders,
although the phase diagram in Fig.\ref{FigPhase} is not because the phase
indicators are not topological invariants. 

First, we introduce randomness into the on-site potential as%
\begin{equation}
V_{n}=V\left( 1+\eta _{n}\zeta \right) ,  \label{Ln}
\end{equation}%
with $\eta _{n}$ being a random variable ranging from $-1$ to $1$, and $%
\zeta $ is the strength of the disorder. The Hamiltonian is modified as 
\begin{equation}
H_{nm}=\kappa _{\text{R}}\delta _{n,m+1}+\kappa _{\text{L}}\delta
_{n,m-1}-(\kappa _{\text{R}}+\kappa _{\text{L}}+V_{n})\delta _{n,m},
\end{equation}%
which we analyze numerically. We show the profile of $S_{n}$ in Fig.\ref%
{FigDisorder}. We observe some features. Although $S_{n}$ is affected
significantly near the edge, the overall skin structure is maintained and
all states remain to be skin states. Namely, the skin states are
robust against disorders even in the nonlinear regime.

Second, we introduce disorders in the hopping%
\begin{eqnarray}
H_{nm} &=&\kappa _{\text{R},n}\delta _{n,m+1}+\kappa _{\text{L,n}}\delta
_{n,m-1}  \notag \\
&&-(\kappa _{\text{R},n}+\kappa _{\text{L,n}}+V_{n})\delta _{n,m},
\end{eqnarray}%
where we have defined%
\begin{equation}
\kappa _{\text{R},n}\equiv \kappa _{\text{R}}\left( 1+\eta _{n}^{\text{R}%
}\zeta \right) ,\quad \kappa _{\text{L,n}}\equiv \kappa _{\text{L}}\left(
1+\eta _{n}^{\text{L}}\zeta \right) ,
\end{equation}%
with $\eta _{n}^{\text{R}}$ and $\eta _{n}^{\text{L}}$ being a random
variable ranging from $-1$ to $1$. The results are shown in Fig.\ref%
{FigDisorderK}. The skin state is largely deformed comparing to the case of
the on-site potential. It is natural because the magnitude of the hopping is
essential for the skin state. However, the overall skin states are preserved
even in the presence of the hopping disorder.

Finally, we introduce disorders in the nonlinearity,%
\begin{equation}
i\frac{d\psi _{n}}{dt}+\kappa _{\text{R}}\psi _{n+1}+\kappa _{\text{L}}\psi
_{n-1}-\left( \kappa _{\text{R}}+\kappa _{\text{L}}\right) \psi _{n}+\xi
_{n}\left\vert \psi _{n}\right\vert ^{2}\psi _{n}=0,
\end{equation}%
where we have defined%
\begin{equation}
\xi _{n}\equiv \xi \left( 1+\eta _{n}\zeta \right) .
\end{equation}%
The results are shown in Fig.\ref{FigDisorderZ}. We find that the skin
state is robust against the nonlinearity disorder.

A comment is in order. The topological number only assures the
robustness of the skin state, and not the robustness of the detailed
structure of the skin state such as the trap-skin, shifted-trap-skin and
embedded-trap-skin states. Actually, they are transformed among them by
disorder effects as shown in Figs.\ref{FigDisorder}, \ref{FigDisorderK} and %
\ref{FigDisorderZ}. This is because the phase indicators are not topological
invariants.

\subsection{Nonlinearity dependence of skin effect}

The strength of the skin effect is suppressed as the nonlinearity increases,
as shown in Fig.\ref{FigSkinDensity}. The amplitude $\left\vert \psi
_{L}\right\vert $ at the right edge is a good signal to estimate it. Its $%
\xi $ dependence is shown in Fig.\ref{FigEdgeZ}, where $\left\vert \psi
_{L}\right\vert $ decreases as the increase of $\xi $. It validates that the
skin state is suppressed by the nonlinear effect. It is due to the fact that
the nonlinear term prefers the stop of the motion, which makes hard for the
initial pulse to hop and form a skin state.

\begin{figure}[t]
\centerline{\includegraphics[width=0.48\textwidth]{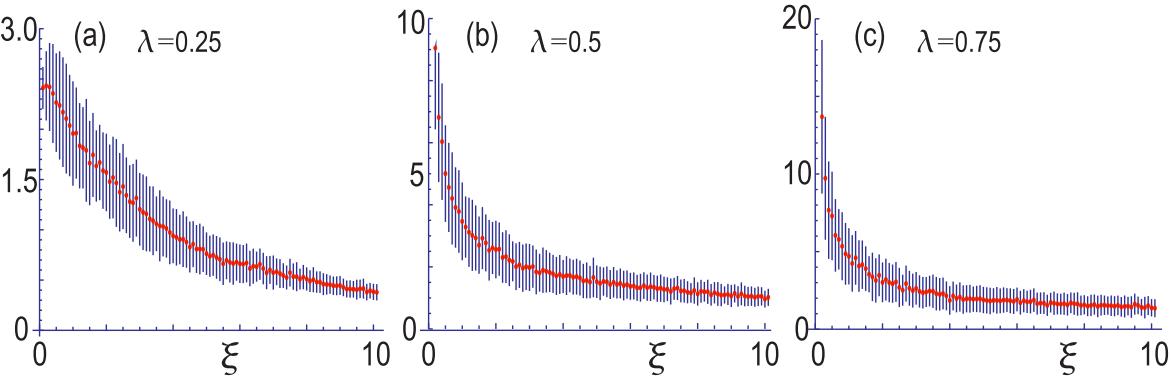}}
\caption{Dependence of the edge amplitude $|\protect\psi _{L}|$ on the
nonlinearity $\protect\xi $ for (a) $\protect\lambda =0.25$, (b) $\protect%
\lambda =0.5$, and (c) $\protect\lambda =0.75$. It decreases as $\protect\xi 
$ increases. We have taken 100 times average on the random on-site
potential. The red dots indicate the mean value, while the blue bars
indicate the standard deviation. We have set $\protect\kappa =1$ and $L=20$.}
\label{FigEdgeZ}
\end{figure}

\section{Two-dimensional model}

\subsection{Nonlinear second-order skin effect}

A two-dimensional generalization of the nonreciprocal discrete nonlinear Schr%
\"{o}dinger equation (\ref{DS}) reads 
\begin{eqnarray}
&&i\frac{d\psi _{n}}{dt}+\kappa _{\text{R}}\psi _{n+x}+\kappa _{\text{L}%
}\psi _{n-x}+\kappa _{\text{U}}\psi _{n+y}+\kappa _{\text{D}}\psi _{n-y} 
\notag \\
&=&\left( \kappa _{\text{R}}+\kappa _{\text{L}}+\kappa _{\text{U}}+\kappa _{%
\text{D}}\right) \psi _{n}-\xi \left\vert \psi _{n}\right\vert ^{2}\psi _{n},
\label{EqB}
\end{eqnarray}%
by introducing the upward and downward hopping amplitudes $\kappa _{\text{U}%
} $ and $\kappa _{\text{D}}$, respectively, in addition to $\kappa _{\text{R}%
}$ and $\kappa _{\text{L}}$. We show the time evolution of the spatial
distribution of the amplitude $\left\vert \psi _{n}\right\vert $ in Fig.\ref%
{FigSoSkin}.

\subsection{Linear model}

We study the linear model with $\xi =0$. In the momentum space, the energy
is given by%
\begin{eqnarray}
E\left( k_{x},k_{y}\right) &=&\kappa _{\text{R}}e^{ik_{x}}+\kappa _{\text{L}%
}e^{-ik_{x}}+\kappa _{\text{U}}e^{ik_{y}}+\kappa _{\text{D}}e^{-ik_{y}} 
\notag \\
&&-\left( \kappa _{\text{R}}+\kappa _{\text{L}}+\kappa _{\text{U}}+\kappa _{%
\text{D}}\right) .
\end{eqnarray}%
When $\left\vert \kappa _{\text{R}}\right\vert <\left\vert \kappa _{\text{L}%
}\right\vert $ and $\left\vert \kappa _{\text{U}}\right\vert <\left\vert
\kappa _{\text{D}}\right\vert $ in the case of a finite square, a
corner-skin state is generated at the left-down corner, whose eigenfunction
of the linear model is given by%
\begin{equation}
\psi _{2n_{x}+1,2n_{y}+1}=\left( -\frac{\kappa _{\text{R}}}{\kappa _{\text{L}%
}}\right) ^{n_{x}}\left( -\frac{\kappa _{\text{U}}}{\kappa _{\text{D}}}%
\right) ^{n_{y}}\psi _{1,1},
\end{equation}%
and otherwise $\psi _{n,m}=0$. All other corner-skin states are obtained in
a similar manner.

\begin{figure*}[t]
\centerline{\includegraphics[width=0.98\textwidth]{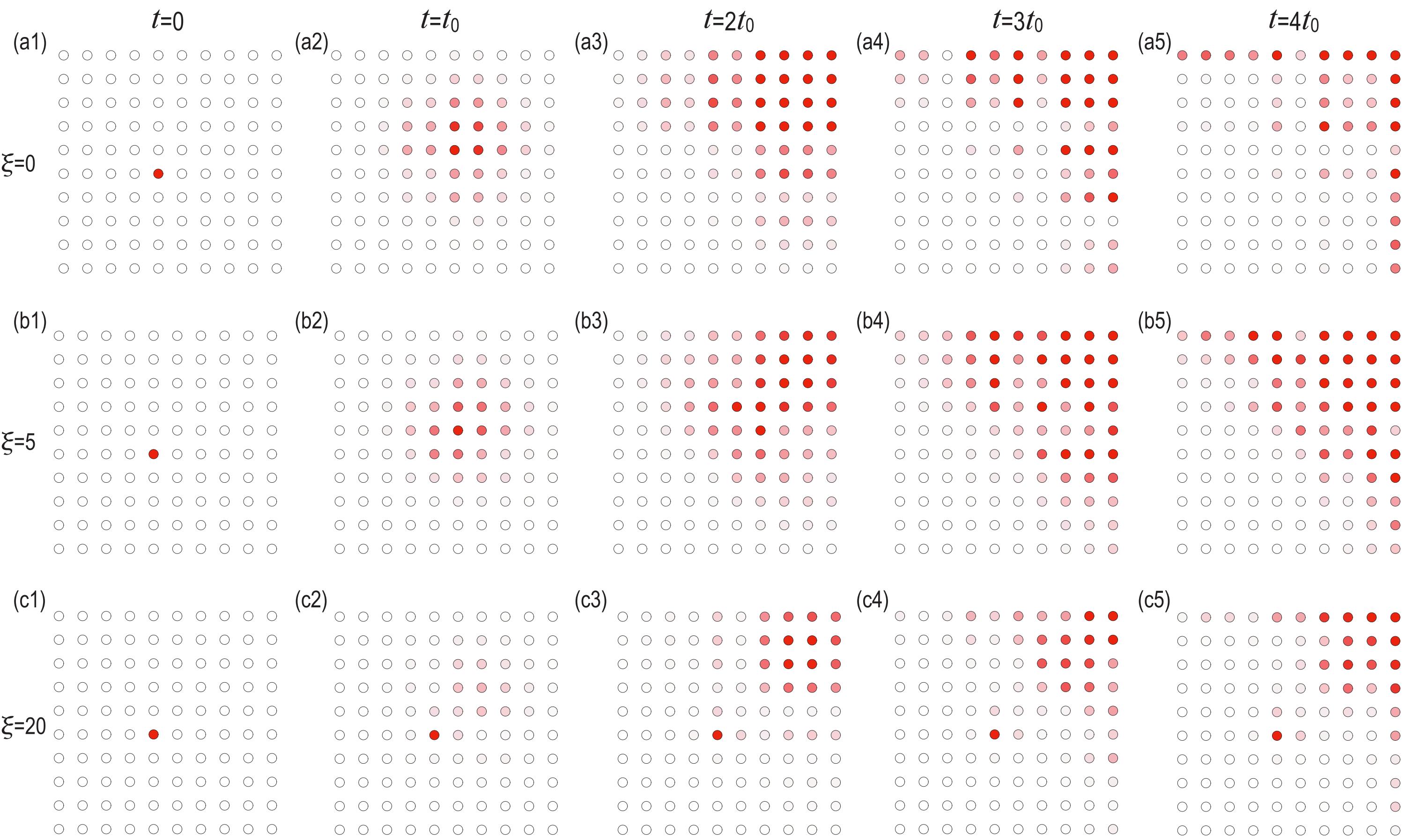}}
\caption{Time evolution of the spatial distribution of the amplitude $|%
\protect\psi _{n}|$ in the two-dimensional model for (a1)$\sim $(a5) $%
\protect\xi =0$, (b1)$\sim $(b5) $\protect\xi =5$, and (c1)$\sim $(c5) $%
\protect\xi =20$, when the quench dynamics starts from the site indicated in
red at $t=0$. The color density indicates the amplitude $|\protect\psi _{n}|$%
. We have set $\protect\kappa _{\text{R}}=\protect\kappa _{\text{U}}=1.5$
and $\protect\kappa _{\text{L}}=\protect\kappa _{\text{D}}=0.5$. We have set 
$\protect\lambda =0.5$, $L=10$. The time step is $t_{0}=1$ and the
simulation time is $T=10$ in units of $1/\protect\kappa $. }
\label{FigSoSkin}
\end{figure*}

\begin{figure}[t]
\centerline{\includegraphics[width=0.48\textwidth]{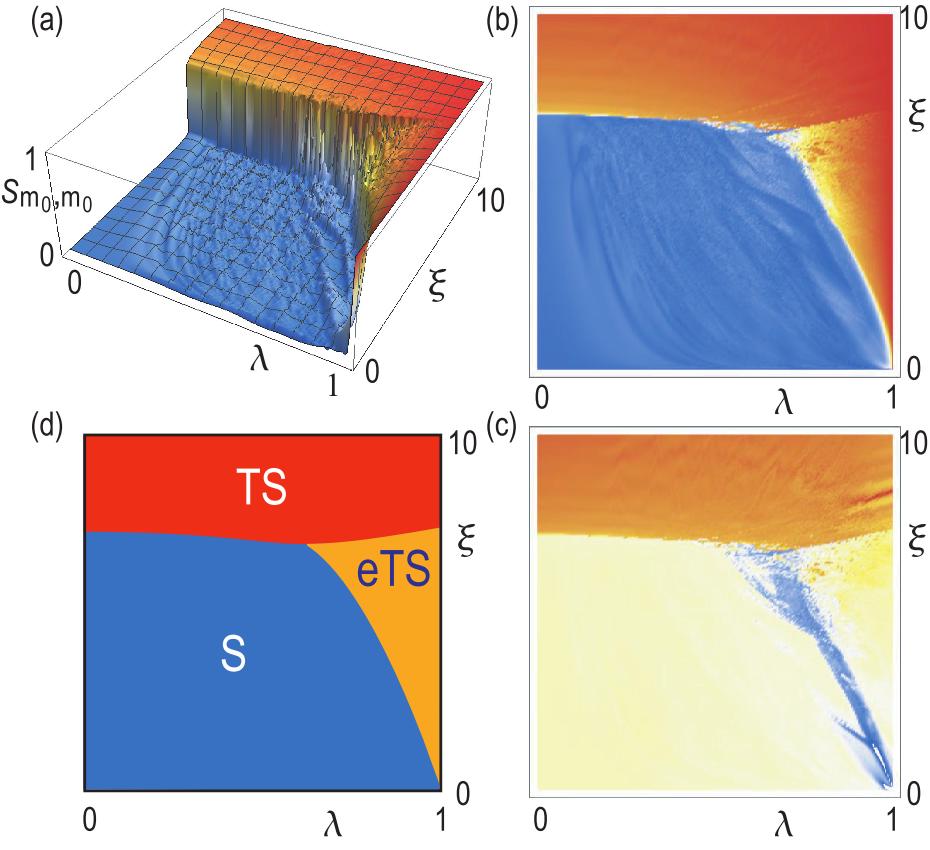}}
\caption{(a) and (b) Time average $S_{m_{0},m_{0}}$ of the amplitude as a
function of the nonreciprocity $\protect\lambda $ and the nonlinearity $%
\protect\xi $ in the two-dimensional model, with (a) being a bird's eye's
view. The red region indicates the trap-skin (TS) phase or the
embedded-trap-skin (eTS) phase where $S_{m_{0},m_{0}}=1$, while the blue
region represents the skin phase where $S_{m_{0},m_{0}}=0$. (c) The time
average of the difference $\Delta S=S_{m_{0},m_{0}}-S_{m_{0},m_{0}+1}$,
which differentiates the eTS phase from the TS phase. (d) Phase diagram in
the $\protect\lambda $-$\protect\xi $ plane. We have set $L=10$ and $T=10$. 
The color palette of (a), (b) and (c) is the same as Fig.3(f).}
\label{FigSoPhase}
\end{figure}

\subsection{Topological number}

The topological number is defined in the linear model by%
\begin{equation}
w_{\mu }=\int_{-\pi }^{\pi }\frac{dk_{\mu }}{2\pi i}\partial _{k_{\mu }}\ln %
\left[ E\left( k_{x},k_{y}\right) -\bar{E}_{\mu }\right] ,
\end{equation}%
where%
\begin{equation}
\bar{E}_{\mu }=\int_{-\pi }^{\pi }\frac{dk_{\mu }}{2\pi }E\left(
k_{x},k_{y}\right)
\end{equation}%
with $\mu =x,y$. We note that $E\left( k_{x},k_{y}\right) -\bar{E}_{\mu }$
only depends on $k_{\mu }$. They are explicitly given by%
\begin{eqnarray}
E\left( k_{x},k_{y}\right) -\bar{E}_{x} &=&\kappa _{\text{R}%
}e^{ik_{x}}+\kappa _{\text{L}}e^{-ik_{x}}, \\
E\left( k_{x},k_{y}\right) -\bar{E}_{y} &=&\kappa _{\text{U}%
}e^{ik_{y}}+\kappa _{\text{D}}e^{-ik_{y}}.
\end{eqnarray}%
The topological numbers are also rewritten as%
\begin{eqnarray}
w_{x} &=&\text{sgn}(\left\vert \psi _{L,L}\right\vert +\left\vert \psi
_{L,1}\right\vert -\left\vert \psi _{1,L}\right\vert -\left\vert \psi
_{1,1}\right\vert ), \\
w_{y} &=&\text{sgn}(\left\vert \psi _{L,L}\right\vert -\left\vert \psi
_{L,1}\right\vert +\left\vert \psi _{1,L}\right\vert -\left\vert \psi
_{1,1}\right\vert ),
\end{eqnarray}%
in a finite square system. If $\kappa _{\text{R}}=\kappa _{\text{U}}$ and $%
\kappa _{\text{L}}=\kappa _{\text{D}}$, they are simply given by%
\begin{equation}
w_{x}=w_{y}=\text{sgn}\left( \left\vert \psi _{L,L}\right\vert -\left\vert
\psi _{1,1}\right\vert \right) ,
\end{equation}%
because we have $\left\vert \psi _{L,1}\right\vert =\left\vert \psi
_{1,L}\right\vert $. It means that there are only two phases, where the skin
states emerge at the right-up corner or the left-down corner.

We have $w_{x}=1$ for $\left\vert \kappa _{\text{L}}\right\vert <\left\vert
\kappa _{\text{R}}\right\vert $ and $w_{x}=-1$ for $\left\vert \kappa _{%
\text{L}}\right\vert >\left\vert \kappa _{\text{R}}\right\vert $, while we
have $w_{y}=1$ for $\left\vert \kappa _{\text{D}}\right\vert <\left\vert
\kappa _{\text{U}}\right\vert $ and $w_{y}=-1$ for $\left\vert \kappa _{%
\text{D}}\right\vert >\left\vert \kappa _{\text{U}}\right\vert $. There are
four topological phases $(w_{x},w_{y})=(\pm 1,\pm 1)$. The corner skin state
emerges at the right-up corner for $(w_{x},w_{y})=\left( 1,1\right) $, at
the right-down corner for $(w_{x},w_{y})=\left( 1,-1\right) $, at the
left-up corner for $(w_{x},w_{y})=\left( -1,1\right) $ and at the left-down
corner for $(w_{x},w_{y})=\left( -1,-1\right) $. These formulas are valid
also in the nonlinear models. The system is always topological unless $%
\kappa _{\text{L}}=\kappa _{\text{R}}$ and $\kappa _{\text{D}}=\kappa _{%
\text{U}}$.

\subsection{Quench dynamics}

We find an analytic solution of the quench dynamics for the linear model (%
\ref{EqB}) with $\xi =0$ as%
\begin{eqnarray}
\psi _{n_{x},n_{y}}\left( t\right) &=&\left( i\sqrt{\frac{\kappa _{\text{L}}%
}{\kappa _{\text{R}}}}\right) ^{n_{x}}\left( i\sqrt{\frac{\kappa _{\text{D}}%
}{\kappa _{\text{U}}}}\right) ^{n_{y}}  \notag \\
&&\times J_{\left\vert n_{x}\right\vert }\left( 2\sqrt{\kappa _{\text{L}%
}\kappa _{\text{R}}}t\right) J_{\left\vert n_{y}\right\vert }\left( 2\sqrt{%
\kappa _{\text{D}}\kappa _{\text{U}}}t\right) ,
\end{eqnarray}%
where it starts from a localized site $m_0=\left( 0,0\right) $ in an
infinite square lattice.

The quench dynamics is numerically studied for the nonlinear model (\ref{EqB}%
), where it starts from a localized site $m_0=\left( L/2,L/2\right) $ in a
finite square lattice. We show the results in Fig.\ref{FigSoSkin}. At the
initial stage, the peak of the amplitude shifts to the up-right direction.
Once it reaches the right-up corner, the state almost remains as it is
although there are fluctuations due to the reflection at the edges. It is a
generalization of the result of the one-dimensional case discussed before.

\subsection{Phase diagram}

We define the time average $S_{m_{0},m_{0}}$ by a straightforward
generalization of Eq.(\ref{Sn}). We show it as a function of the
nonreciprocity $\lambda $ and the nonlinearity $\xi $ in Figs.\ref%
{FigSoPhase}(a) and (b), where Fig.\ref{FigSoPhase}(a) is a bird's eye's
view. The blue region indicates the skin phase. We also plot $\Delta S\equiv
S_{m_{0},m_{0}}-S_{m_{0},m_{0}+1}$ in Fig.\ref{FigSoPhase}(c), which
differentiates the trap-skin and the embedded-trap-skin states. As in the
case of the one-dimensional model, there are skin, trap-skin and
embedded-trap-skin states in the phase diagram. On the other hand, there is
no shifted-trap-skin state in the two-dimensional model. It may be due to
the fact that there is a large degree of freedom to hop in the
two-dimensional model in contrast to the one-dimensional model, which makes
harder for the initial pulse to form a shifted-trap-skin state but results
in just an ordinary skin state. We thus obtain a phase diagram as in Fig.\ref%
{FigSoPhase}(d).

\section{Discussions}

We have studied models in which nonreciprocity and nonlinearity coexist.
Skin states emerge irrespective of the nonlinearity, which implies that the
system is always topological. Additionally, we have found four typically
different states generated by the nonlinearity effect in the one-dimensional
model. They are the trap-skin, shifted-trap-skin and embedded-trap-skin
states, forming four phases.

Two comments are in order. First, skin states are also formed at dislocations%
\cite{Bha,Schind}. It is an interesting problem to study nonlinear skin
effect in systems with dislocations. Second, nonlinear non-Hermitian skin
states are studied recently in the frequency space\cite{YuceSkin}. However,
the four phases above mentioned are not found in this work. In addition,
there is no generalization to nonlinear higher-order skin states. The quench
method reveals a rich phase diagram of the nonlinear nonreciprocal system.

The author is very much grateful to S. Iwamoto and N. Nagaosa for helpful
discussions on the subject. This work is supported by the Grants-in-Aid for
Scientific Research from MEXT KAKENHI (Grants No. JP17K05490 and No.
JP18H03676). This work is also supported by CREST, JST (JPMJCR16F1 and
JPMJCR20T2).

\end{document}